\documentstyle[preprint,aps]{revtex}

\begin{document}
\draft
\preprint{}
\title{Spectral functions in doped transition metal oxides}
\author{D.D. Sarma and S.R. Barman}
\address{Solid State and Structural Chemistry Unit, Indian Institute of
Science, Bangalore 560012, INDIA}
\author{H. Kajueter and G. Kotliar}
\address{
Serin Physics Laboratory, Rutgers University, Piscataway, NJ 08855 USA}
\date{\today}
\maketitle
\begin{abstract}
We present experimental photoemission and inverse photoemission
spectra of SrTiO$_{3- \delta}$ representing electron doped $d^0$
systems. Photoemission spectra in presence of electron doping exhibit
prominent features arising from electron correlation effects, while
the inverse photoemssion spectra are dominated by spectral features
explainable within single-particle approaches. We show that such a
spectral evolution in chemically doped correlated systems is not
compatible with expectations based on Hubbard or any other similar
model. We present a new theoretical approach taking into account the
inhomogeneity of the {\it real} system which gives qualitatively
different results compared to standard {\it homogeneous} models and
is in quantitative agreement with experiments.
\end{abstract}

\newpage
        Following the discovery of high T$_c$ oxides \cite{BM}, there
has been a resurgence of interest in 
doped transition metal oxides. Recent investigations have revealed
some very unusual physics in these systems \cite{Tokura,Fuji,Fuji1}.
La$_x$Sr$_{1-x}$TiO$_3$ \cite{Tokura} is a system containing Ti $3d$
electrons  where the carrier concentration can be varied in a
controlled fashioned.  At one end, LaTiO$_3$ is a Mott-Hubbard
insulator with a $3d^1$ configuration, while at the other end of the
solid solution, SrTiO$_3$ is a band insulator with a $3d^0$
configuration. Thus, varying $x$ tunes the electron configuration
continuously between 0 and 1. For large values of $x$, the transport
and magnetic properties \cite{Tokura} as well as the experimentally
observed spectral weights behave as a correlated doped Mott
insulator.  For small values of $x$ ($\approx$ 0), the system
represents small amount of electron doping in a $3d^0$ configuration,
and thus, it should behave as a {\it doped band insulator} and is
expected to show little correlation effects.  This expectation is
also consistent with transport and magnetic properties of such
compounds \cite{Tokura,transport} which exhibit an essentially
free-electron like behavior.

Interestingly however, ultra-violet photoemission spectra (UPS) of
La$_x$Sr$_{1-x}$TiO$_3$ with small $x$ (= 0.1 and 0.2) exhibit two
features:  a coherent low energy feature and a  very prominent
incoherent feature which is in fact found to be {\it more intense}
than the coherent feature\cite{Fuji1}. The  observation of a large
fraction of incoherent to coherent spectral weight suggests the
presence of very strong correlation effects, in contrast to transport
and magnetic properties, as well as to the theoretical expectations
of an uncorrelated behavior close to the $d^0$ configuration.
Coherent  features at low energies and incoherent features at around
1.4 eV are also seen in  systems with one d  electron per unit
cell\cite{Fuji}.  The evolution of the redistribution of spectral
weight as a function of interaction strength in the half filled
system can be accounted by zero temperature large d calculations
based on the Hubbard model \cite{gabi}.  The doping dependence of
these features, however, has not been accounted within this
framework.

In order to clarify these important issues, we have reinvestigated
the electronic structure of lightly doped $d^0$ systems both
experimentally and theoretically.  We supplement the already existing
UPS of La$_x$Sr$_{1-x}$TiO$_3$ with those of SrTiO$_{3- \delta}$,
where the oxygen deficiency ($\delta$) introduces electrons into the
stoichiometric $d^0$ system.  We  have also  carried out inverse
photoemission experiments in these systems.  Since we expect that the
strength of the incoherent features is more visible in the unoccupied
part of the spectrum in systems which have a small number of $d$
electrons, this technique is ideal for settling the importance of
correlations here.  The present results conclusively establish that
the observed spectral function is  inconsistent with the {\it
homogeneous} Hubbard model  which is normally used in the
interpretation of these kinds of experiments. We present a new
theoretical treatment of inhomogeneity (arising from disorder in the
real system) within a Hubbard-like model to establish qualitatively
new descriptions for the evolution of spectral functions 
in chemically substituted systems, in very good 
agreement with experimental results. Whereas substituted compounds
have been extensively investigated in recent times, the role of
disorder has almost always been neglected; thus, our findings have
important implications for a large number of compounds.

Polycrystalline ingots of oxygen deficient SrTiO$_{3- \delta}$ with
$\delta$ = 0.0, 0.08 and 0.28 were prepared by melt quenching
required amounts of Sr$_2$TiO$_4$, TiO and TiO$_2$ in an 
inert gas arc furnace.  
Oxygen contents were determined
thermogravimetrically.  The sample surfaces were cleaned by {\it
in-situ} scraping using an alumina file at liquid nitrogen
temperature to avoid surface degradation. UPS and bremsstrahlung
isochromat (BI) spectra were obtained with 0.1 and 0.8 eV resolutions
respectively, in a commercial VSW spectrometer \cite{Sarma1}.
Self-consistent semi-relativistic {\it ab-initio} calculations with
the linearized muffin-tin orbital method within the atomic sphere
approximation (LMTO-ASA) for SrTiO$_{3-\delta}$ with oxygen vacancies
were carried out by considering a cubic supercell consisting of eight
formula units of SrTiO$_3$ with empty spheres at oxygen vacancy sites
and 216 $k$-points in the irreducible part of the Brillouin zone.

In Fig. 1 we show the experimental He {\scriptsize I} UPS of
SrTiO$_{3- \delta}$ for $\delta$ = 0.0, 0.08 and 0.28 near the Fermi
energy (E$_F$) region;
the inset shows a typical spectrum over a wider energy range, being
very similar for all the compositions with the intense doublet
feature at about 7 eV 
arising from primarily oxygen
$2p$ derived states. 
Expectedly, the wide band gap material, SrTiO$_3$ shows no intensity
at E$_F$.  On introducing oxygen vacancies ($\delta$ $>$ 0), 
it is clear that two spectral
features emerge in the near E$_F$ region, one appearing at E$_F$ and
another at  about 1.3 eV, in striking
similarity to those of La$_x$Sr$_{1-x}$TiO$_3$ \cite{Fuji1}.  While
the feature at E$_F$ is due to electron doping and the consequent
movement of 
 the Fermi energy  $E_F$ 
into the conduction band as confirmed by
our band structure results on SrTiO$_{3- \delta}$, the calculations
do not have any counterpart of the experimentally observed 1.3 eV
feature (see later in the text).  Furthermore, this higher binding
energy feature in Fig. 1 (and also in La$_x$Sr$_{1-x}$TiO$_3$
\cite{Fuji1}) is similar to the incoherent feature observed in
LaTiO$_3$ \cite{Fuji} which is an end member of the solid solution
La$_x$Sr$_{1-x}$TiO$_3$. These observations along with the absence of
this feature in single-particle calculations, would strongly suggest
correlation effects being responsible for the origin of this peak in
these lightly doped electron systems.

In Fig. 2 we show BI spectra of SrTiO$_3$ and SrTiO$_{2.72}$ compared
to the theoretically calculated BI spectrum of SrTiO$_3$ obtained
from band structure calculations and calculated matrix elements
\cite{Sarma1}. The calculated spectrum provides a 
good description of the experimentally observed one for SrTiO$_3$ within
the first 10 eV of E$_F$; the mismatch at higher energies arises from
the well-known limitation of linearized band structure methods. This
agreement is consistent with the fact that SrTiO$_3$ with a
$d^0$ configuration is primarily a band insulator with correlation
effects playing no major role. However, it is interesting to note
that the BI spectrum of SrTiO$_{2.72}$ is almost identical to that of
SrTiO$_3$, except for a small shift ($\approx$ 0.5 eV) of the peaks
and some overall broadening of the spectral features. The 
shift is consistent with electron doping of the system and thus,
is explainable within a rigid shift of E$_F$ with respect to the
one-electron results for SrTiO$_3$ with no perceptible signature of
any correlation induced features in contrast to the occupied spectral
features (Fig. 1).  We, however, cannot rule out the presence of weak
intensity features corresponding to the upper Hubbard band which may
be considerably broadened if it coexists in the region of 4-9 eV with
extended O and Sr states which have a much larger intensity.

The above results of the presence of correlation induced feature in
the occupied parts and the simultaneous absence of any such feature in
the unoccupied parts, are very surprising.
It is well known that for few electrons in the band
(such as Ce and U containing systems),
correlation induced features appear most prominently in the
unoccupied spectral function, whereas in Ni based systems with few
holes in the band, the correlation effects are prominent only in the
occupied parts.
We demonstrate explicitly that  the evolution of
the photoemission and inverse photoemission spectra as a function of
doping cannot be accounted by one band  or a multiband   homogeneous model
by  incorporating
self-energy corrections to the calculated Ti $d$ density of states
(DOS) in SrTiO$_{3- \delta}$ ($\delta$ = 0.125) within a second order
perturbative treatment of correlation\cite{Treglia}.  The results for
$U$ = 0 and 4.6 eV are shown in Fig. 3. 
It is clear from this figure
that a correlation induced satellite appears outside the one-electron
($U$ = 0) bandwidth in the unoccupied parts with increasing $U$ and
no such distinct feature can be seen in the occupied parts. Similar 
results are also found within exact diagonalization studies of 
the Hubbard model\cite{cluster} and in the solution of the Hubbard 
in infinite dimensions, while an {\it exactly opposite trend} 
is observed in the experimental results in Figs. 1 and 2.

In order to understand this surprising result, we first note that the
Hubbard model represents a homogeneous system, whereas in real
systems electron (or hole) doping is brought about by chemical
substitution; {\it this invariably introduces disorder or
inhomogeneity at some length scale}. 
  In this paper  we propose that
it is essential to use 
 a spatially  {\it inhomogeneous} picture
to model  these systems.
 The basic idea is that in these systems the
electrons experience different local environments.  The angle
integrated  photoemission spectrum, as a local probe, captures a
suitable average of the Green's function over the local environments.
In principle the degree of inhomogeneity can vary between two limits.
If the system is disordered on a microscopic scale
 (termed microscopic inhomogeneity here),
but homogeneous
on a macroscopic scale, i.e. the impurities do not form clusters, a
model of uncorrelated disorder is appropriate. 
If there is a
tendency of the defects to form clusters, this may reinforce the
natural tendency of strongly correlated systems to phase separate as
pointed out in ref\cite{Emery} and
a  model of macroscopic inhomogeneities is more suitable.  Recent
developments in the treatment of correlated electrons by using a
mean field theory that becomes exact in the limit of infinite spatial
coordination \cite{rev} allows us to make these ideas more
quantitative.  We study the evolution of photoemission spectra in
these two well defined limits.  We expect that the physical situation
is somewhere in between. 
We stress that in both limits we
find the photoemission spectrum as a function of $x$ to be very
different from the predictions of the Hubbard model and very similar
to what is observed experimentally.

In the microscopically inhomogeneous situation we describe
SrTiO$_{3-\delta}$ or La$_{x}$Sr$_{1-x}$TiO$_3$ using a  Hubbard
model with diagonal disorder.  The Hamiltonian is given by: 
\begin{displaymath}
H = - \sum_{(ij)} t_{ij\sigma} C^{+}_{i\sigma}C_{j\sigma} + \sum_{i}
(\epsilon_{i} - \mu) C^{+}_{i\sigma}C_{i\sigma} + \sum_{i}
n_{i\uparrow} n_{i\downarrow} U_{i}
\end{displaymath}
To get a bounded density of states and a well defined limit as the
lattice coordination gets large, we take the hopping matrix elements
$t_{ij}$ to be defined on a Bethe lattice and scale them as
$\frac{t}{\sqrt{d}}$.  
The full
bandwidth $2 D = 4 t$, is estimated to be 2.5 eV from our band 
structure calculations.
The $\epsilon_{i}$ are random variables which take  the values
$\epsilon_{A}$ with probability $P_{A}$ and $\epsilon_{B}$ with
probability $P_{B}$; $P_A$ and $P_B$ are given by the composition, as
$2 \delta$ and $(1- 2 \delta)$ in SrTiO$_{3-\delta}$ and $x$ and
$(1-x)$ in La$_x$Sr$_{1-x}$TiO$_3$ respectively.  
Supercell calculations for SrTiO$_{2.875}$ and SrTiO$_{2.75}$ showed
that several bands are pulled down compared to SrTiO$_3$ due to the
presence of vacancies by about 1.5 - 3.0 eV which provides an
estimate for $(\epsilon_B - \epsilon_A)$. We use $(\epsilon_B -
\epsilon_A)$ $\sim 2.4$ eV; however, the qualitative features of the
calculations do not depend on this exact value. $U_A = U_B = $ 4.6 eV
was estimated from earlier spectroscopic analysis.  The chemical
potential $\mu$ is chosen so that the number of $d$ electrons is
equal to $x$ in La$_x$Sr$_{1-x}$TiO$_3$ or $2\delta$ in
SrTiO$_{3-\delta}$.

The mean field theory which is exact  in the  $d \rightarrow \infty$
limit determines the local  Green's functions at the A and B sites as
the local $f$ electron Green's function of an  Anderson impurity model
\begin{eqnarray}
H & = & \sum_{k} \epsilon_{k} a^{+}_{k\sigma} a_{k\sigma} +
U f^{+}_{\uparrow} f_{\uparrow}
f^{+}_{\downarrow} f_{\downarrow}\nonumber \\
 & & +  \sum_{k} V_{k}(a^{+}_{k\sigma}f_{\sigma} + f^{+}_{\sigma}
a_{ko})
 +  \epsilon_{f} f^{+}_{\sigma} f_{\sigma}
\label{anderson}
\end{eqnarray}
Here the $f$ level position  $\epsilon_{f}$ is given by  $E_{A}-\mu$ or
$E_{B}-\mu$  respectively.  The hybridization function
$\Delta(i \omega_n)$
\begin{equation}
 \Delta(i \omega_n)= 
\sum_{k}
\frac{V^{2}_{k}}{(i\omega_{n} - \epsilon_{k})}
\label{hyb}
\end{equation}
has to be determined self consistently from the local Green's
functions via the equation:
\begin{displaymath}
\nonumber
\Delta(i\omega_n) = t^2\ G_{av}(i \omega_n)
=t^2\ ( P_{A}\ G_{A}(i\omega_n) + P_{B}\ G_{B}(i\omega_n))
\end{displaymath}

It is to be noticed that the A and B sites are connected by the same
hybridization function which depends on the local physics at both
sites. Thus we have a coupled system of equations.  To solve the mean
field equations we used the recent extension of the IPT
method\cite{IPT} to extract real frequency information. On the
imaginary axis identical results are obtained with the exact
diagonalization algorithm of Caffarel and Krauth\cite{caffarel}.

We now turn to the case where the inhomogeneities
occur on a macroscopic scale.
In this
scenario the density  of electrons is inhomogeneous on a more
macroscopic scale.  A fraction $f_1$ is in an insulating phase with
the  local concentration of the Mott insulating phase  $n_{ins}=1$
and a fraction $1-f_1$  with the metallic concentration $n_{met}$.
The total density $n={f_1}n_{ins}+(1- {f_1})n_{met}$, and the
measured Green's function is given by
\begin{equation}
\label{separate}
\bar{G} = {f_1} G_{n_{ins}} + (1-{f_1}) G_{n_{met}}
\end{equation}
$G_{n_{ins}}$ and $G_{n_{met}}$ can again be calculated  using the
mean field theory now in the respective homogeneous phases.  To
explore this scenario we took $n_{met}=.1$ (which corresponds to a
chemical potential $\mu=-.78$) and varied the density $n$ by varying
the volume fraction ${f_1}$ in equation (\ref{separate}).  The
spectral function in these two scenarios  for a density close to
$n=.2$ are shown in Fig. 4 a and b including a finite broadening to
model the finite experimental resolutions. Fig. 4 a and b are in very
good agreement with the experimental spectra (Figs. 1 and 2); this is
more striking so, when we keep in mind that we expect the real
system to be somewhere in between the two limiting cases
theoretically considered here. For example, the calculation correctly
predicts the existence of an intense incoherent feature at about 1.3
eV arising from the local Green's function for site A with a band-like
coherent feature near $E_F$ due to the local Green's function for site
B. This is in perfect agreement with the experimentally observed
features in the photoemission spectra of SrTiO$_{3-\delta}$ (Fig. 1)
and La$_x$Sr$_{1-x}$TiO$_3$ \cite{Fuji1}.  Moreover, it is clear from
the calculated results that unoccupied parts of the spectral function
will be entirely dominated by coherent band-like features arising
from site B as observed in Fig. 2. Here we once again draw
attention to the fact that these results are in sharp contrast to
those expected on the basis of Hubbard and similar homogeneous models
(see Fig. 3). While there have been past attempts to deal with
substoichiometric correlated systems \cite{refs}, effects of disorder
inevitably present in such systems have not been addressed so far.

To summarize,  in this paper we studied 
high-energy spectroscopic data of compounds with close
to 3$d^0$ configuration.
We presented explicit calculations of the spectral function of
correlated electrons at different dopings in an inhomogeneous picture.
We considered two cases, in one the inhomogeneity is present only on
a microscopic scale while in the other the system was assumed to be
macroscopically inhomogeneous. Both models give similar angle integrated
spectra which therefore 
cannot  determine the length scale
over which the system is inhomogeneous. 
The  spectra of the inhomogeneous models,
however are qualitatively different from that of
a homogeneous system,  conclusively establishing the need to go beyond
the standard theories to describe spectral functions in a class of doped
transition metal compounds.

FIGURES:

Fig. 1 He {\scriptsize I} photoemission of SrTiO$_{3- \delta}$ for
three values of $\delta$.

Fig. 2 Bremsstrahlung isochromat spectra of SrTiO$_3$ and
SrTiO$_{2.72}$.

Fig. 3  2nd order perturbation applied to the DOS obtained from LMTO
supercell calculations for SrTiO$_{2.875}$.

Fig. 4 Spectral functions of the disordered Hubbard model for denisty
n=.2 in the case of (a) microscopic and (b) macroscopic
inhomogeneities. 
The separate contributions from the A and B sites to the total
are also shown. 
The occupied parts of the spectral functions are
expanded for clarity, as shown in the figures.
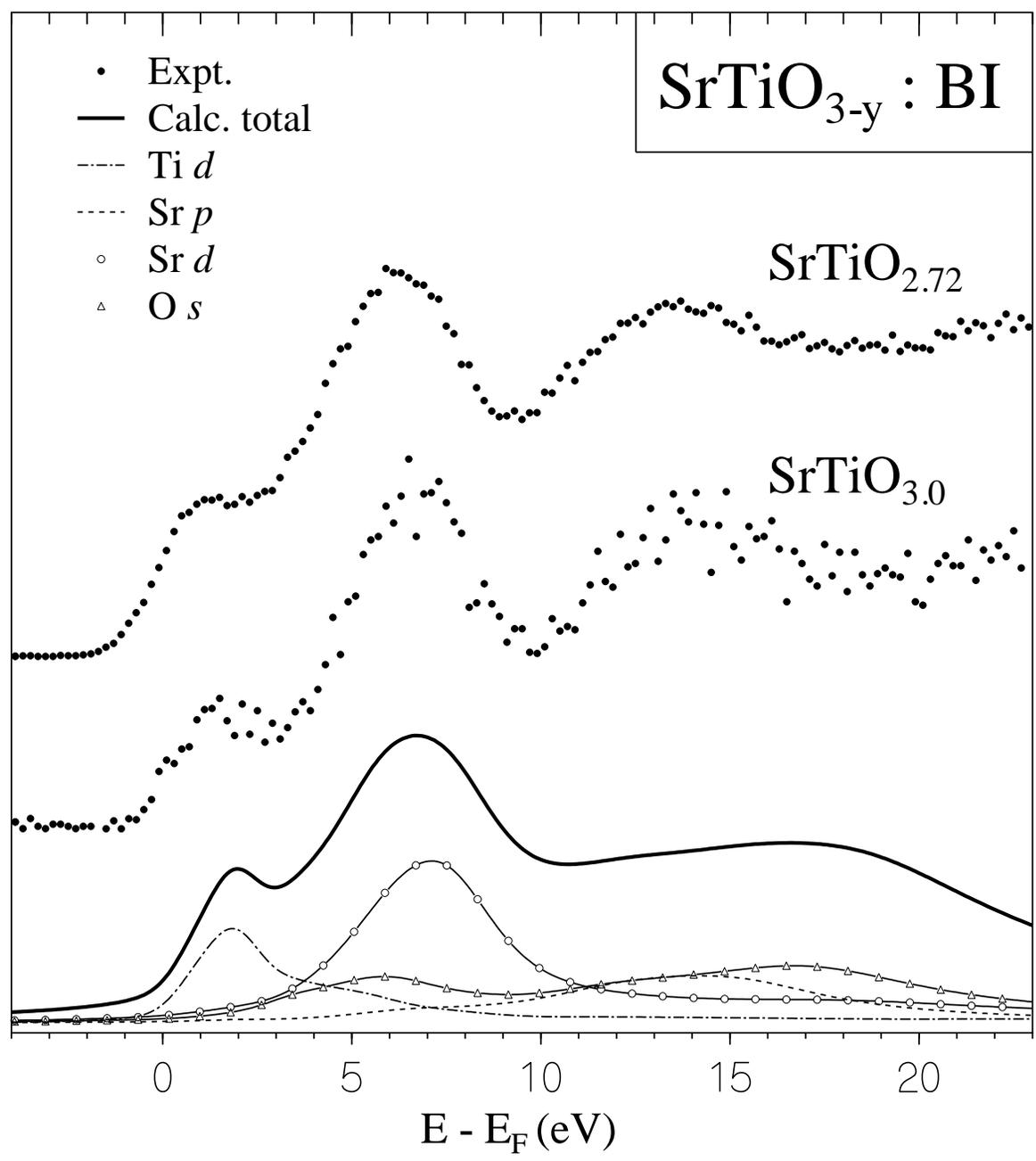
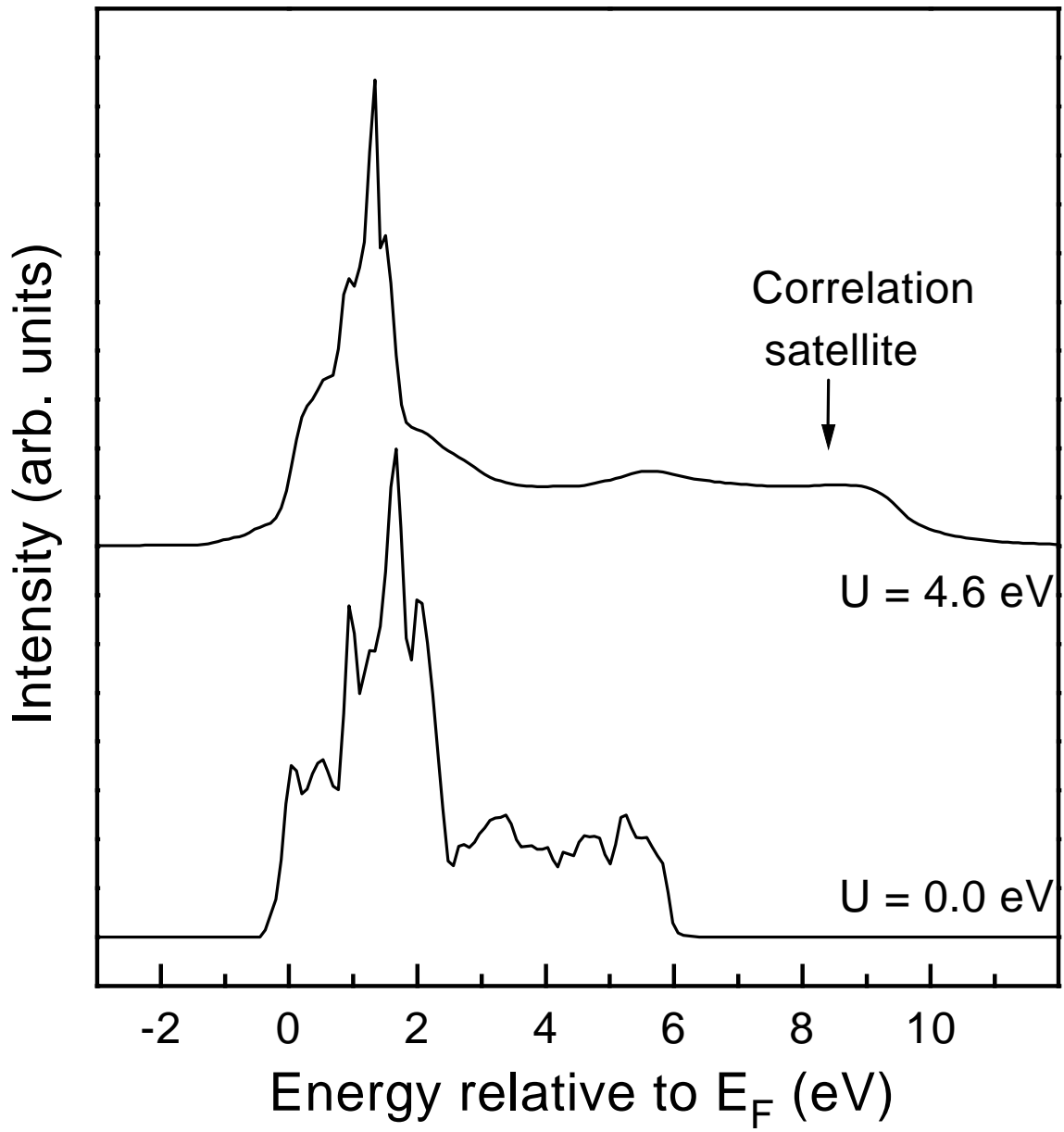
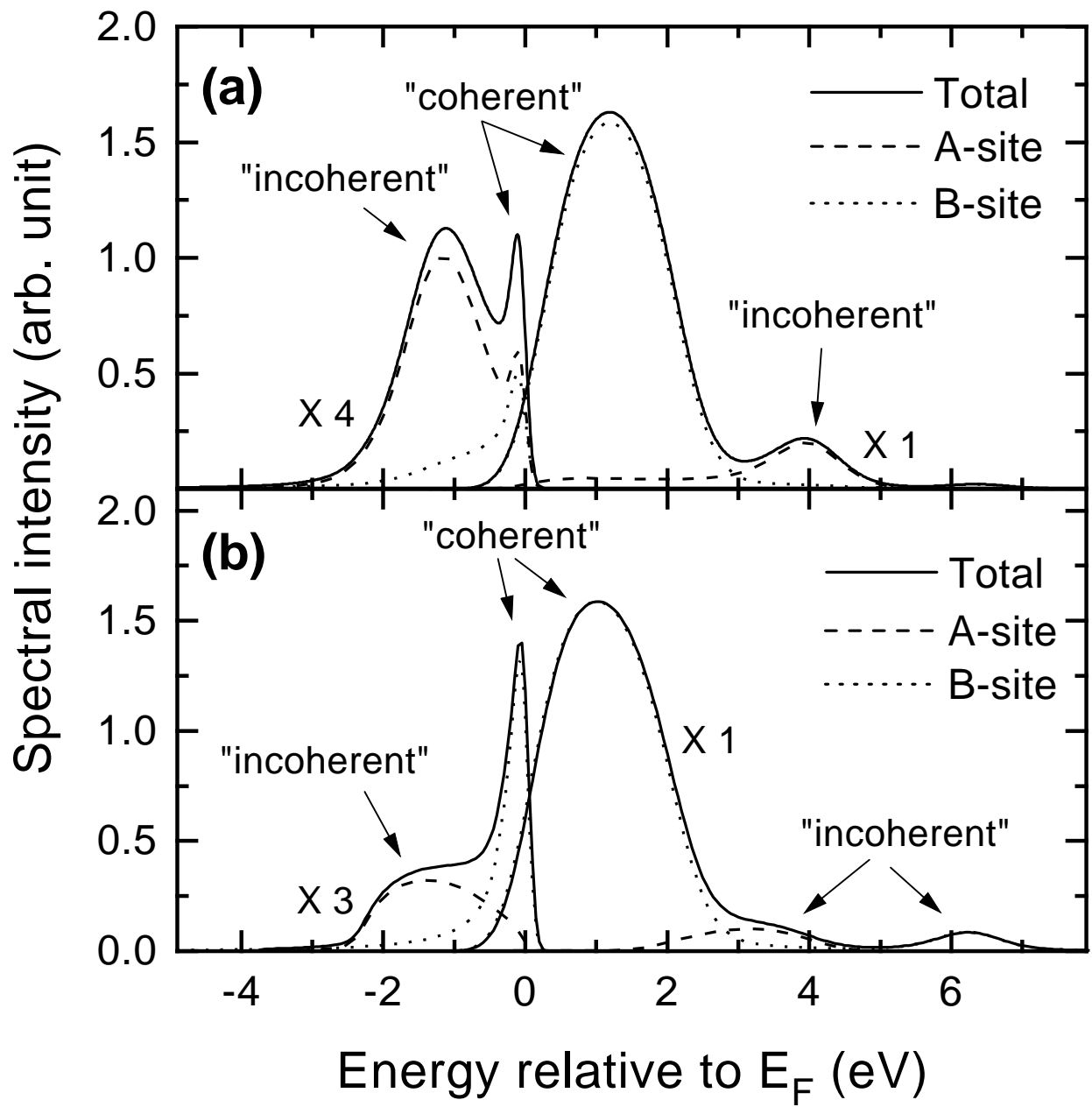
\end{document}